\documentclass{sigchi-ext}
\usepackage[T1]{fontenc}
\usepackage{textcomp}
\usepackage[scaled=.92]{helvet} % for proper fonts
\usepackage{graphicx} % for EPS use the graphics package instead
\usepackage{balance}  % for useful for balancing the last columns
\usepackage{booktabs} % for pretty table rules
\usepackage{ccicons}  % for Creative Commons citation icons
\usepackage{ragged2e} % for tighter hyphenation

\def\plaintitle{SIGCHI Extended Abstracts Sample File: Note Initial
  Caps} 
\def\emptyauthor{}
\def\plainkeywords{Projection-based Augmented Reality; Dementia; Elderly People; Long-term Memory; Mental Therapy; Health Care}

\title{A Projection-based Augmented Reality \protect\\ for Elderly People with Dementia}

\numberofauthors{6}
\author{%
  \alignauthor{%
    \textbf{Hyocheol Ro}\\
    \affaddr{Media System Lab} \\
    \affaddr{Yonsei University} \\
    \affaddr{Seoul, Republic of Korea} \\
    \email{hyocheol.ro@msl.yonsei.ac.kr} } \vfil 
  \alignauthor{%
    \textbf{Yoon Jung Park}\\
    \affaddr{Media System Lab} \\
    \affaddr{Yonsei University}\\
    \affaddr{Seoul, Republic of Korea}\\
    \email{yoonjung.park@msl.yonsei.ac.kr} } \vfil 
  \alignauthor{%
    \textbf{Tack-Don Han}\\
    \affaddr{Media System Lab} \\
    \affaddr{Yonsei University} \\
    \affaddr{Seoul, Republic of Korea} \\
    \email{hantack@msl.yonsei.ac.kr} }
}

\definecolor{linkColor}{RGB}{6,125,233}
\hypersetup{%
  pdftitle={\plaintitle},
  pdfauthor={\emptyauthor},
  pdfkeywords={\plainkeywords},
  bookmarksnumbered,
  pdfstartview={FitH},
  colorlinks,
  citecolor=black,
  filecolor=black,
  linkcolor=black,
  urlcolor=linkColor,
  breaklinks=true,
}

\begin{document}

%% For the camera ready, use the commands provided by the ACM in the Permission Release Form.
%\CopyrightYear{2007}
%\setcopyright{rightsretained}
%\conferenceinfo{WOODSTOCK}{'97 El Paso, Texas USA}
%\isbn{0-12345-67-8/90/01}
%\doi{http://dx.doi.org/10.1145/2858036.2858119}
%% Then override the default copyright message with the \acmcopyright command.
%\copyrightinfo{\acmcopyright}

\maketitle

% Uncomment to disable hyphenation (not recommended)
% https://twitter.com/anjirokhan/status/546046683331973120
\RaggedRight{} 

% Do not change the page size or page settings.
\begin{abstract}
As aging societies grow, researchers are actively studying care systems concerning the life and diseases of the elderly. Among these diseases, dementia makes it difficult to maintain daily life due to the degradation of cognitive functioning, memory, and reasoning, as well as the ability to perform actions. Moreover, dementia does not have a perfect cure, though therapy and care can slow its onset and provide patients with physical and mental support. In this paper, we developed a projection-based augmented reality system robot that can cover 360 degrees of space. We also propose an application that supports continuous monitoring of dementia patients to address the difficulties they face in daily life. The system is also designed to provide therapy applications, such as entertainment and spatial art, to provide mental care aids for the patients.
\end{abstract}

% ACM Classfication



% Author Keywords

% Print the classficiation codes
% Please use the 2012 Classifiers and see this link to embed them in the Tex: \url{https://dl.acm.org/ccs/ccs_flat.cfm}

\section{Introduction}
The aging society continues to grow as quality of life and medical technology improve. Recently, dementia was found to be the third leading cause of death in the elderly, after heart disease and cancer. Dementia makes it difficult for the elderly to maintain daily life due to the degradation of cognitive functioning, memory, and reasoning, as well as the ability to perform actions. Causes of the illness vary with changes occurring in the brain. Elderly dementia patients are expected to reach 46.8 million worldwide by 2015 and double every 20 years \cite{prince2015global}. However, there is currently still no complete cure for dementia. 
Current systems slow the onset of the disease or assist people suffering from it. To reduce memory loss and support various therapies, studies are being conducted using augmented reality (AR) and virtual reality (VR) that can access a wide range of media and content. Research is also being conducted to address neurodevelopmental disorders and improve attention by providing visual experiences such as 360 degree panoramic video to the elderly through VR headsets. However, VR headsets can cause dizziness (motion sickness and digital point-of-view strain) and have interfaces difficult for the elderly to use \cite{ohyama2007autonomic}. 
\begin{marginfigure}[-10pc]
  \begin{minipage}{\marginparwidth}
    \centering
    \includegraphics[width=0.9\marginparwidth]{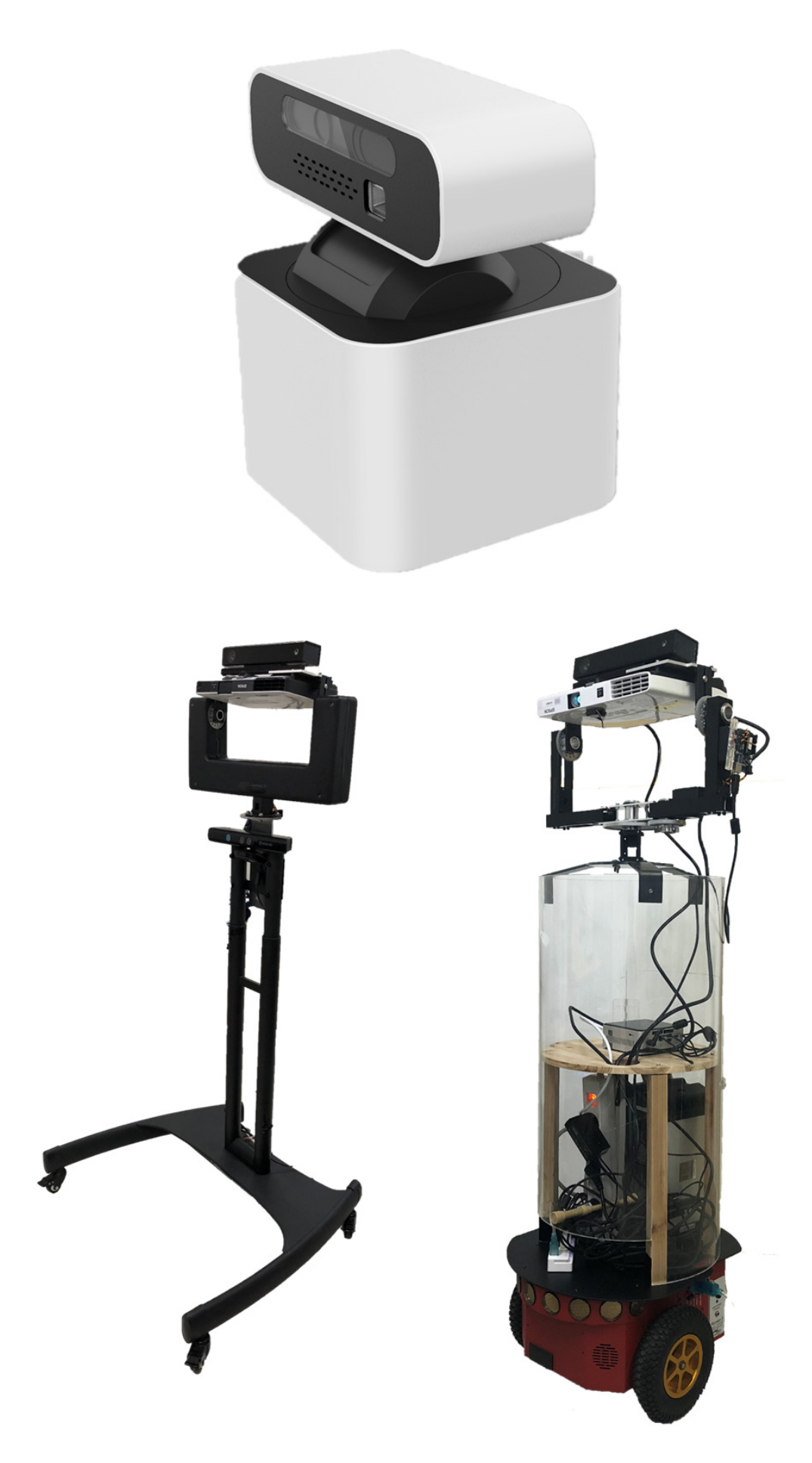}
    \caption{Our Projection AR System. (top) Portable System. (bottom left) Movable System. (bottom right) Mobile Robot System.}
    ~\label{fig:system}
  \end{minipage}
\end{marginfigure}
Various studies have also used AR headsets to address the problems of VR headsets; the former improves provides visual information to the real world while improving tangibility and inducing little dizziness, nausea, and the like. Recent studies have been conducted on therapeutic systems to stimulate short-term memory and spatial memory using HoloLens \cite{aruanno2017hololens}. However, HoloLens has a narrow viewing angle and requires the device to be worn continuously, inconveniencing the user. Projection-based AR systems that can address these problems and simultaneously transmits information to many people and provides virtual information directly to the real world through projection to deliver high realism. Research is being conducted on therapeutic care using interactive artwork that supports user interaction with projection AR systems. 
This paper proposes a system that can construct an AR environment for all the spaces in which the user exists. This will provide the role of life support, care, and therapy to the elderly and patients living alone or in nursing homes to slow the onset of diseases and provide convenience to their daily life. Accordingly, the proposed system supports mental care aids such as spatial art therapy and memory photo reproduction, in addition to physical support and memory assistance activities. This system can therefore slow the progression of the disease of dementia patients and improve their quality of life through therapy-based mental care.

% \begin{figure*}
%   \centering
%   \includegraphics[width=1\textwidth]{figures/cats}
%   \caption{In this image, the map maximizes use of space. You can make
%     figures as wide as you need, up to a maximum of the full width of
%     both columns. Note that \LaTeX\ tends to render large figures on a
%     dedicated page. Image: \ccbynd~ayman on Flickr.}
%     ~\label{fig:cats}
% \end{figure*}

\section{System Configuration}
Our projection AR system is implemented in various forms as shown in Fig.\ref{fig:system}. The common point of each system is that they can analyze a space and track the users and provide the same service even if they are put into an unfamiliar environment. The existing researches that provided the immersive media room form through the projection was not possible until the whole system was built in the specific space. On the other hand, no separate installation is required to use our system. Therefore, our system can be put into any environment immediately at any time. 3D map reconstruction using SLAM algorithm was implemented to detect the plane to provide the projection. To do this, we installed an RGB-D camera on a pan-tilt system capable of rotating 360 degrees, and the system automatically rotates and collects peripheral space information. 
%The collected 3D data is reconstructed in the system to find the optimal plane. 
RANdom SAmple Consensus (RANSAC)-based plane detection works to obtain candidate planes that can be projected. The applications all support speech recognition and (depth) touch interfaces, and the feedback to them is also made up of voice and visual projection.

\section{Applications}
\subsubsection{Monitoring}
\begin{marginfigure}[-15pc]
  \begin{minipage}{\marginparwidth}
    \centering
    \includegraphics[width=0.9\marginparwidth]{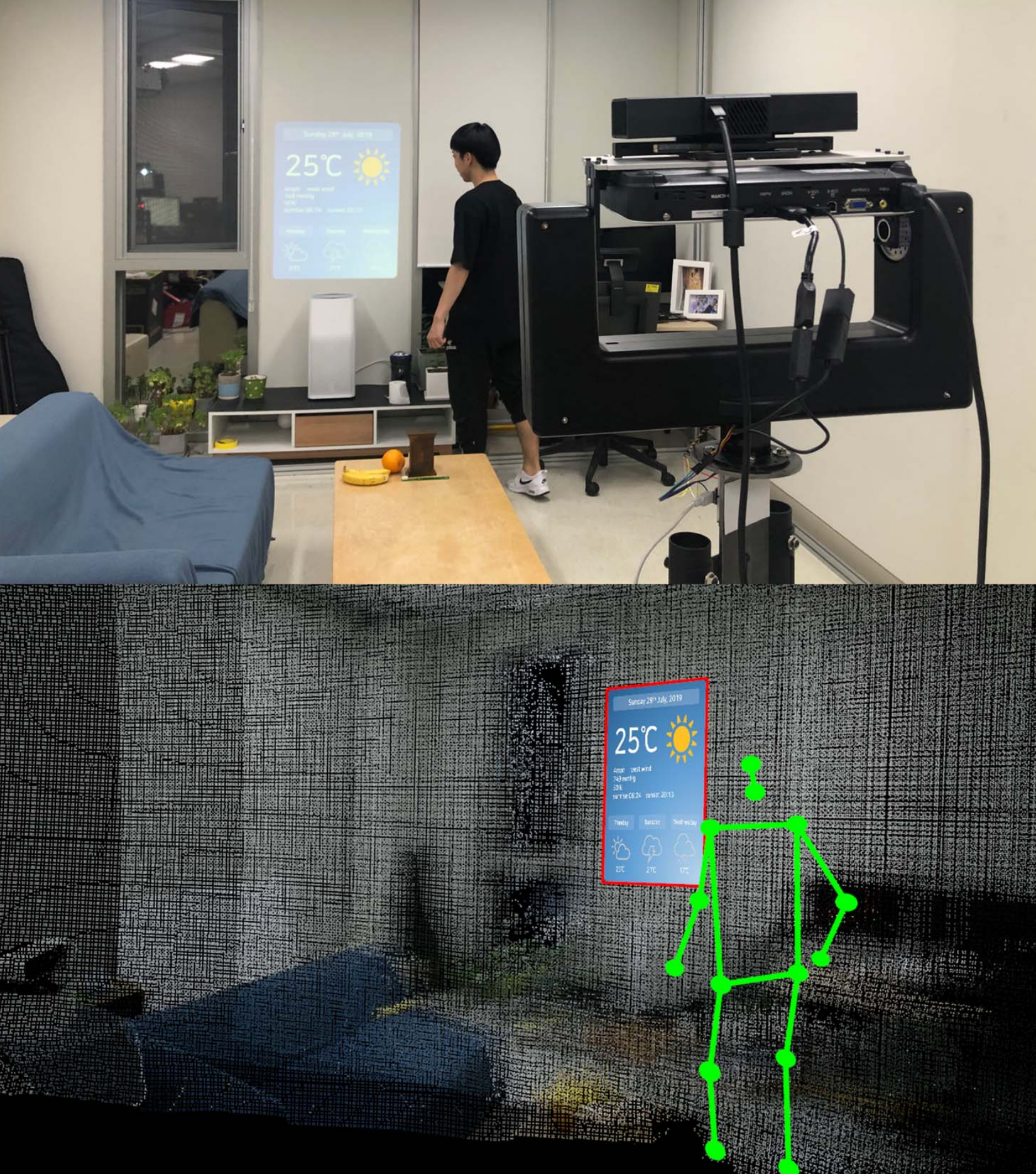}
    \caption{Monitoring Application.}
    ~\label{fig:monitoring}
  \end{minipage}
\end{marginfigure}

After the entire environment is constructed, the proposed system continues to track the user's position through the tracking camera as shown in Fig.\ref{fig:monitoring} . The user's position is expressed in 3D coordinates in the space through the 3D camera, and is used to augment the information through projection when the user approaches the plane on which the content is displayed. The system is also used to monitor abnormal situations of the elderly as well as falls and emergencies. For elderly users living alone, emergency situations often worsen because they cannot request medical support. Accordingly, the system rotates the pan tilt system, continuously monitors the position and status of the user, and provides information to the connected family members to prepare for an emergency situation.
% \ref{fig:marginfig} % 참고하라할때 사용할 것
\begin{marginfigure}[0pc]
  \begin{minipage}{\marginparwidth}
    \centering
    \includegraphics[width=0.9\marginparwidth]{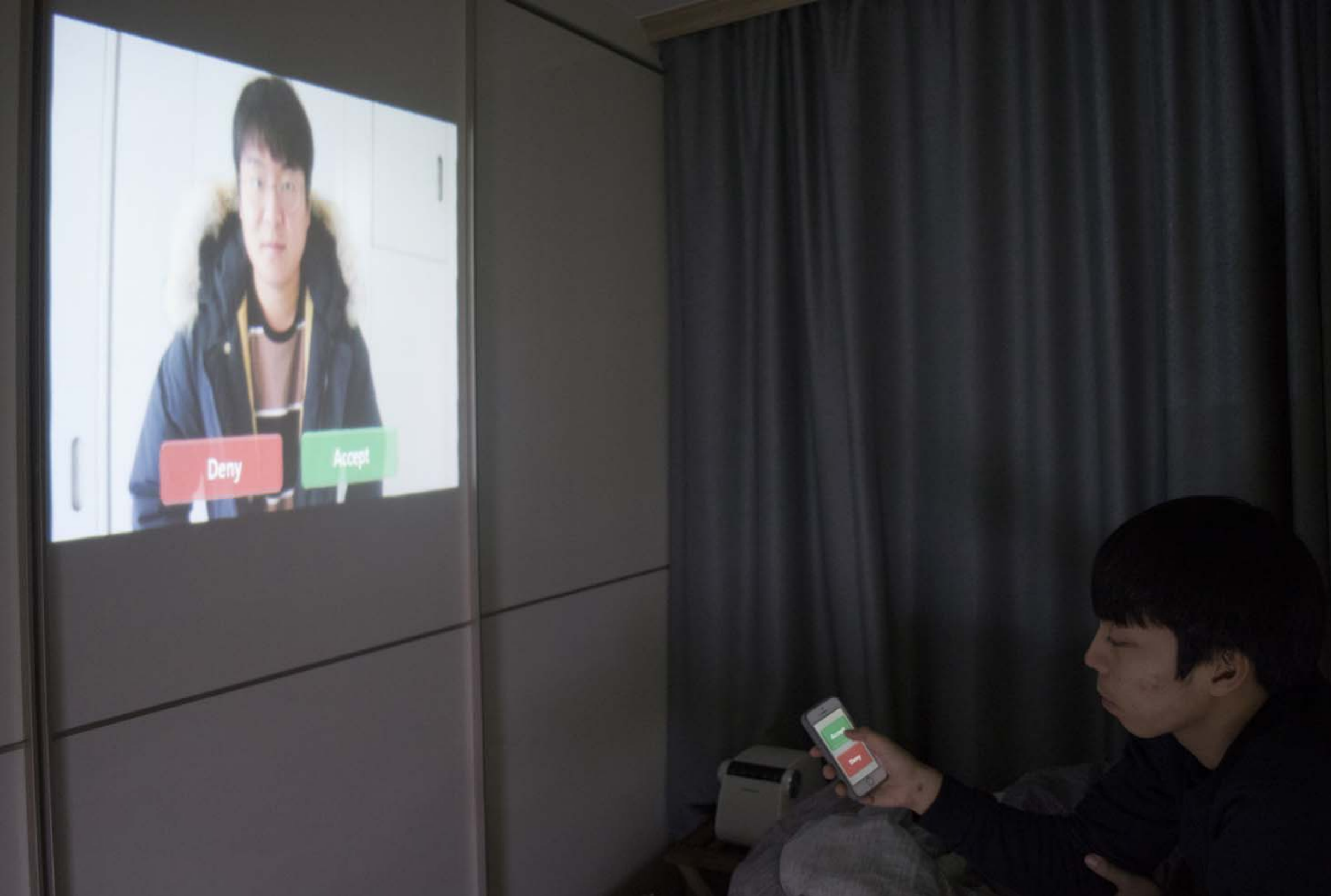}
    \caption{Internet of Things Application.}
    ~\label{fig:iot}
  \end{minipage}
\end{marginfigure}
\begin{marginfigure}[00pc]
  \begin{minipage}{\marginparwidth}
    \centering
    \includegraphics[width=0.9\marginparwidth]{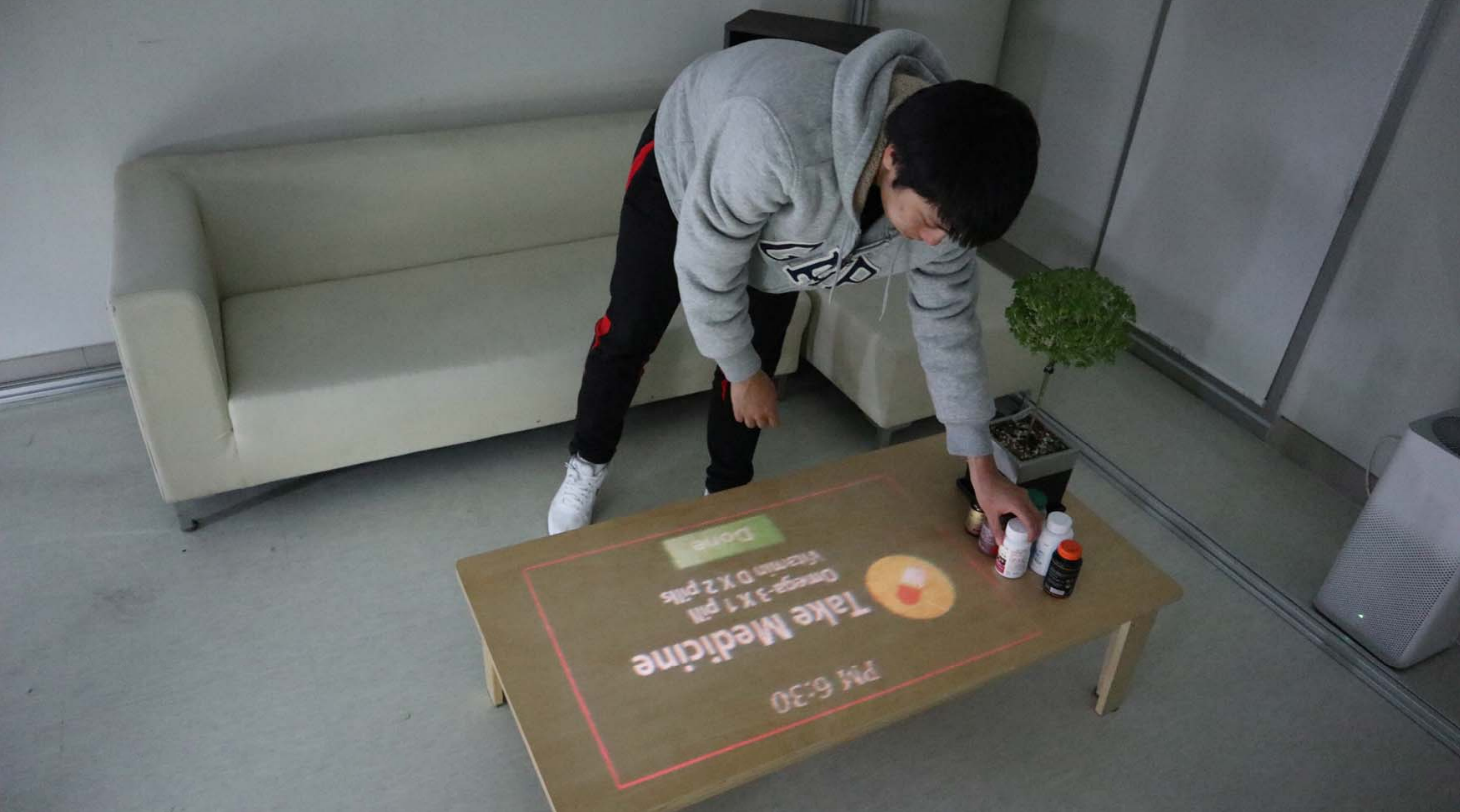}
    \caption{Personal Assistant Application.}
    ~\label{fig:personal-assistant}
  \end{minipage}
\end{marginfigure}
\subsubsection{Internet of Things / User Identification}
Elderly people who suffer from long-term memory problems brought on by diseases such as dementia can lose the ability to recognize faces \cite{kawagoe2017face}. This phenomenon is particularly pronounced in the ability to identify others in a short time. In other words, even if the patient does not forget the other person due to dementia symptoms, they must often spend long periods of time remembering who the person is. The proposed system can provide a solution to this problem: an IoT application that is able to identify the user as shown in Fig.\ref{fig:iot}. Assuming that a guest visiting the home of the elderly user presses the doorbell, the user is displayed a screen of the IoT door on a certain plane through the projection AR \cite{yang2018care}. 
%We added a deep learning module that identifies users before distributing the screen provided here as AR information. 
If the user on the screen is registered in advance, then that person's name and brief information are provided to the elderly user. Meanwhile, if the user is not registered, the result "Unknown" is provided to alert the elderly user. 

\subsubsection{Personal Assistant}
Next, we introduce a personal assistant application that notifies the user of the time to take their medicine \cite{yang2018care}. In addition to mental illnesses such as dementia, elderly people also suffer from chronic diseases such as hypertension and diabetes. The prognosis of these chronic diseases varies greatly with how they are managed after diagnosis. However, medication adherence (taking medication properly according to instructions from a specialist) for elderly patients is often reduced as they may easily forget their dosage or when to take the medicine. 
%As a result, misuse or abuse of the medicine leads to various side effects. 
For this purpose, various mobile applications that notify the time to take the medicine have been introduced. However, because the elderly are generally not familiar with using mobile devices such as smartphones, these methods are also limited. In contrast, our proposed dosing notification system provides dosing time intuitively and effectively through auditory and visual effects as shown in Fig.\ref{fig:personal-assistant}. Depending on the situation, a projection is displayed near the elderly user or in the direction they are looking, and after taking the medicine, the alarm is stopped.

\subsubsection{Media Services}
\begin{marginfigure}[0pc]
  \begin{minipage}{\marginparwidth}
    \centering
    \includegraphics[width=0.9\marginparwidth]{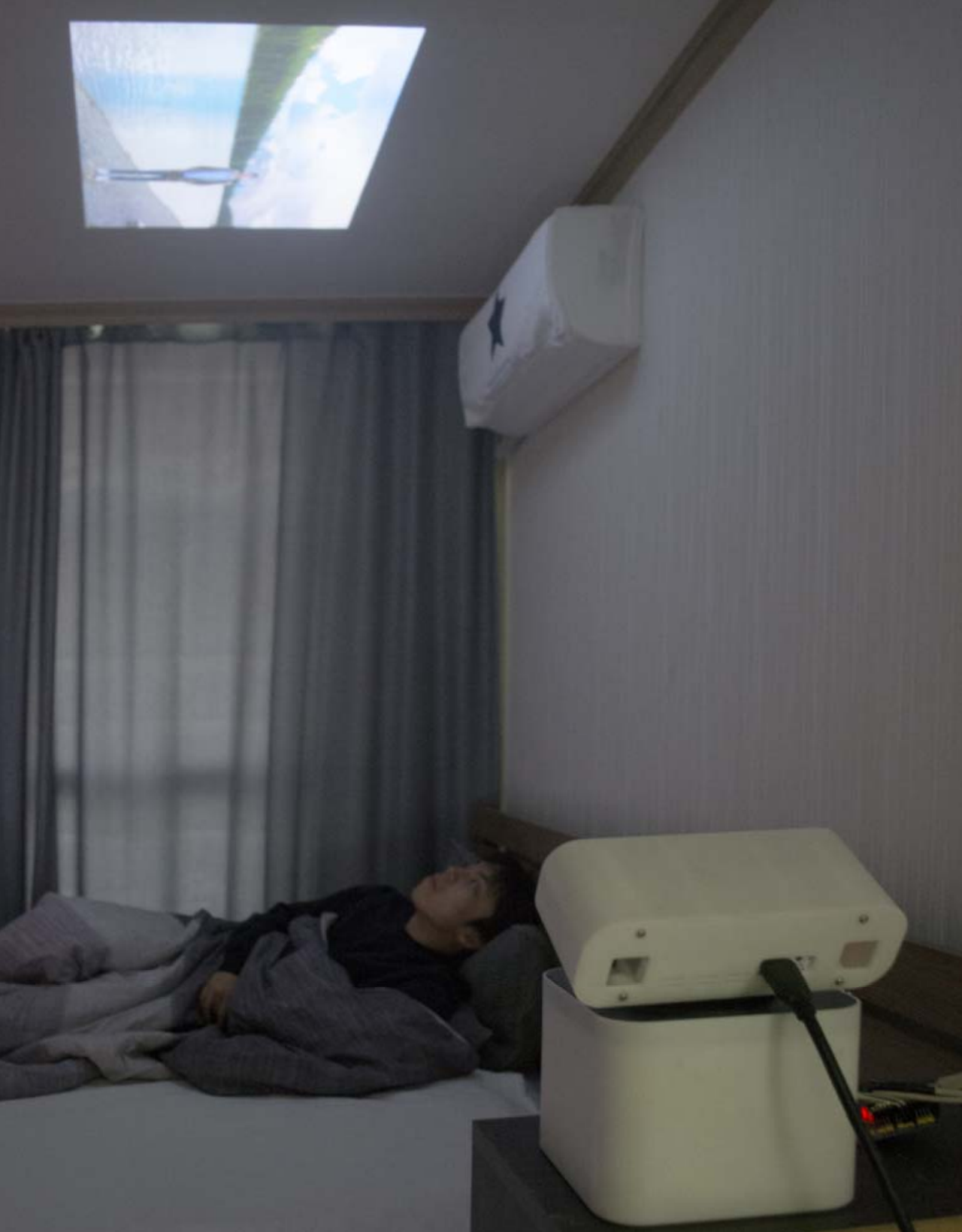}
    \caption{Media Service (Memories) Application.}
    ~\label{fig:media-services}
  \end{minipage}
\end{marginfigure}

According to a statistical survey \cite{pang2015older}, media literacy is significantly reduced for the elderly living alone who have little contact with younger people. This indicates that the information gap (digital divide) between the elderly and the younger generation is becoming more severe in aged societies. Particularly for younger generations, the proportion of media utilization (photo and video viewing, etc.) through mobile devices increases daily, while elderly cannot use the same services even if they have access to the same device \cite{yang2018care}. 
Therefore, rather than teaching the user how to operate the mobile device, we implemented a method that directly links the system with media services to provide them to the elderly. Fig.\ref{fig:media-services} is an example of an application in which the screen is linked to memories of iOS. The media service automatically edits the photos and videos taken by the user and provides it to them in video form \cite{yang2018care}; this service is linked to help improve the media utilization of the elderly user. Among methods to address the symptoms of dementia, one involves using a photograph that can remind the patient of certain people or places. Such media services can be provided to alleviate the symptoms of dementia.
% \ref{fig:marginfig} % 참고하라할때 사용할 것
\begin{marginfigure}[0pc]
  \begin{minipage}{\marginparwidth}
    \centering
    \includegraphics[width=0.9\marginparwidth]{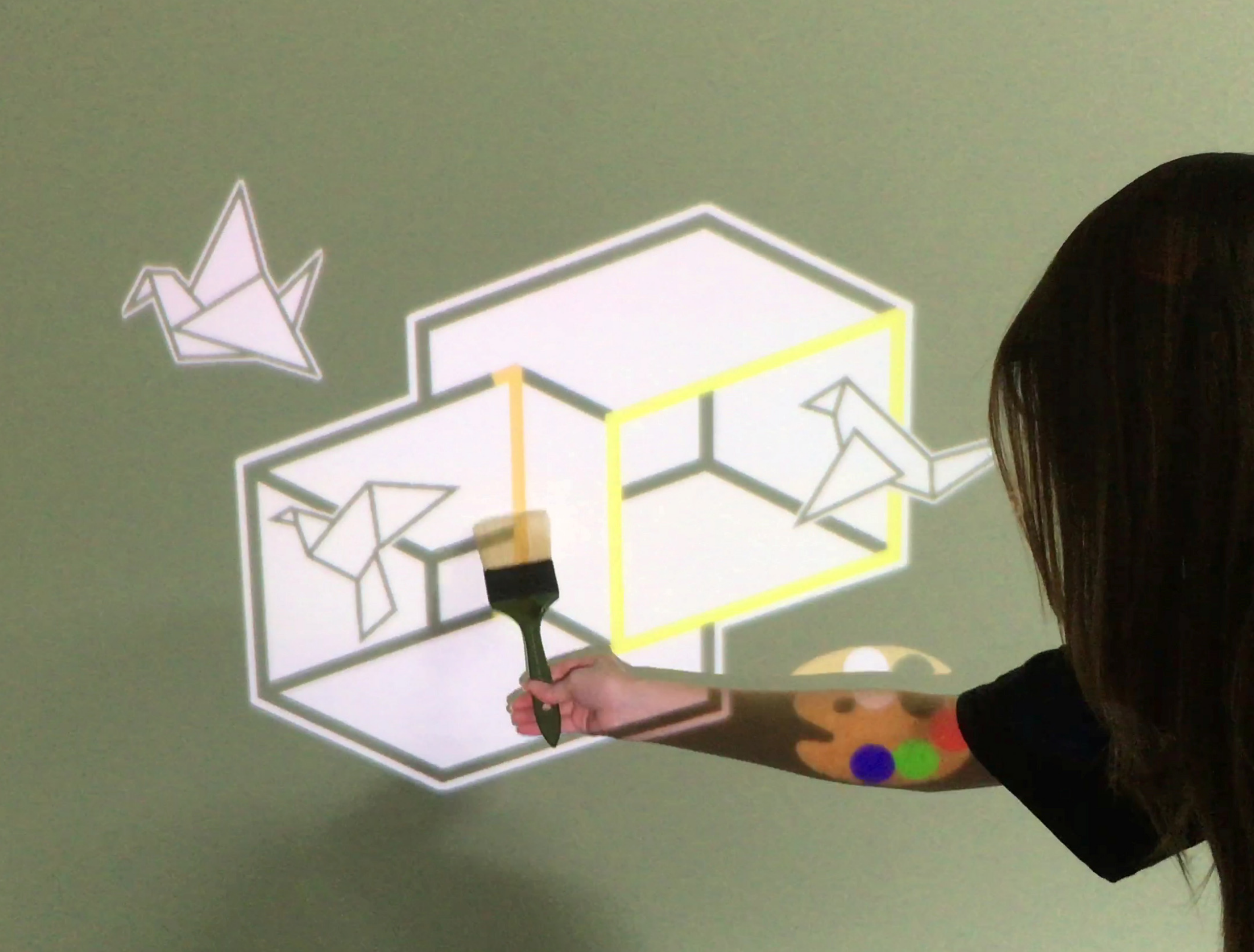}
    \caption{Painting Spatial Art Application.}
    ~\label{fig:spatial-art1}
  \end{minipage}
\end{marginfigure}
\begin{marginfigure}[0pc]
  \begin{minipage}{\marginparwidth}
    \centering
    \includegraphics[width=0.9\marginparwidth]{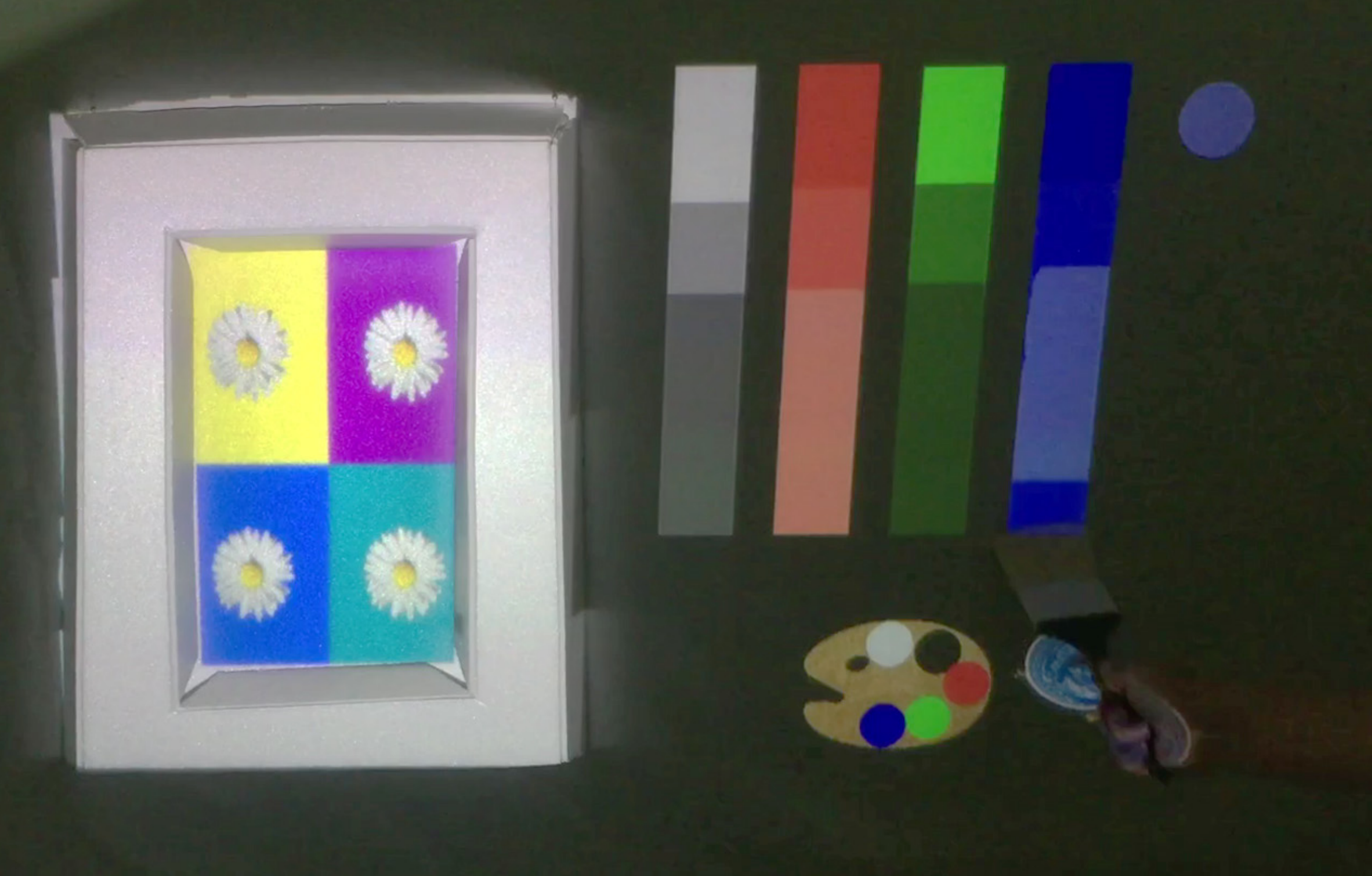}
    \caption{Color Mixing Spatial Art Application.}
    ~\label{fig:spatial-art2}
  \end{minipage}
\end{marginfigure}
\subsubsection{Spatial Art}
A scenario that supports the mental therapy of users suffering from dementia, Spatial Touch is a spatial art application in which the user interacts with the interface to combine various colors and paint or draw as shown in Fig.\ref{fig:spatial-art1} and \ref{fig:spatial-art2}. Through AR spatial art based on visual projection, patients can improve their freedom of expression and imagination, while the use of therapy also assists with their mental care. By interacting through a paintbrush, a metaphor for the same action in the real world, the improvement of the user's physical ability is supported. In real world settings with no separate touch sensors, this method can support intuitive and natural interactions through spatial touch interactions. The user is presented an example picture through the projection, and the user can trace the picture or paint based on it using the brush, the interaction metaphor. 

\section{Conclusion}
%At a time when the world is turning into an aging society, the number of dementia patients continues to increase, but the system to care for them is very scarce. 
In this paper, we propose a variety of applications for care of patients with dementia using the projection AR system. It provides the functions to support mental and physical difficulties. It also supports media services and therapies that can reduce emotional lability. Our AR platform provides proper information and intuitive interactions. Therefore, it can provide various services to dementia patients effectively. In future works, we will explore the future possibilities of projection AR and dementia care by demonstrating these applications directly to patients and receiving feedback.

\bibliographystyle{SIGCHI-Reference-Format}
\bibliography{extended-abstract}

\end{document}